\documentclass[conference]{IEEEtran}

\usepackage{url}
\usepackage{ifthen}
\usepackage[noadjust,space]{cite}
\usepackage[cmex10]{amsmath} 

\usepackage[a-3b]{pdfx}
\hypersetup{
	colorlinks=true,
	allcolors=blue
}

\usepackage{bm}
\usepackage{bbm}
\usepackage{amsfonts}
\usepackage{xspace}
\usepackage{mathtools}

\usepackage[inline]{enumitem}

\usepackage{boxedminipage}
\usepackage{adjustbox}

\usepackage[nameinlink,capitalize]{cleveref}
\newadjustboxenv{balgo}{max width={\linewidth}, innerenv={boxedminipage}{\linewidth}}
\newenvironment{boxedalgo}{\begin{center}\begin{balgo}}{\end{balgo}\end{center}}

\newtheorem{theorem}{Theorem}
\newtheorem{informalthm}{Informal Theorem}
\newtheorem{corollary}{Corollary}

\newtheorem{lemma}{Lemma}

\newtheorem{definition}{Definition}

\newtheorem{remark}{Remark}

\newcommand{\bbC}{\ensuremath{\mathbb{C}}\xspace}

\newcommand{\bbN}{\ensuremath{\mathbb{N}}\xspace}

\newcommand{\bbR}{\ensuremath{\mathbb{R}}\xspace}

\newcommand{\bbZ}{\ensuremath{\mathbb{Z}}\xspace}

\newcommand{\cA}{\ensuremath{\mathcal{A}}\xspace}

\newcommand{\cF}{\ensuremath{\mathcal{F}}\xspace}

\DeclareMathOperator{\polylog}{polylog}
\DeclareMathOperator{\poly}{poly}

\newcommand{\zo}{\ensuremath{{\{0,1\}}}\xspace}

\newcommand{\negl}{\ensuremath{\mathsf{negl}}\xspace}
\newcommand{\eps}{\ensuremath{\varepsilon}}

\newcommand{\defeq}[0]{\ensuremath{{\;\vcentcolon=\;}}\xspace}

\newcommand{\GF}[1]{\ensuremath{\mathbb{GF}}\xspace}

\newcommand{\ie}{\text{i.e.}\xspace}
\newcommand{\etal}{\text{et al.}\xspace}

\newcommand{\whp}{\text{w.h.p.}\xspace}

\newcommand{\eg}{\text{e.g.}\xspace}

\newcommand{\ala}{\text{\`{a} la}\xspace}

\newcommand{\superscript}[1]{\ensuremath{^{\mbox{\tiny{\textit{#1}}}}}\xspace}
\def \th {\superscript{th}}     

\newcommand{\pred}[1]{\ensuremath{\mathsf{#1}}\xspace}

\newcommand{\Gen}{\ensuremath{\mathsf{Gen}}\xspace}
\newcommand{\Enc}{\ensuremath{\mathsf{Enc}}\xspace}
\newcommand{\Dec}{\ensuremath{\mathsf{Dec}}\xspace}
\newcommand{\sk}{\ensuremath{\mathsf{sk}}\xspace}
\DeclareMathOperator{\dist}{dist}
\newcommand{\HAM}{\ensuremath{\mathsf{HAM}}\xspace}
\newcommand{\ED}{\ensuremath{\mathsf{ED}}\xspace}
\newcommand{\fin}{\pred{f}}

\newcommand{\Encf}{\ensuremath{\Enc_{\fin}}\xspace}
\newcommand{\Decf}{\ensuremath{\Dec_{\fin}}\xspace}
\newcommand{\Genf}{\ensuremath{\Gen_{\fin}}\xspace}

\newcommand{\LDC}{\ensuremath{\mathsf{LDC}}\xspace}

\newcommand{\privLDCgame}{\ensuremath{\mathtt{priv}\text{-}\mathtt{LDC}\text{-}\mathtt{Game}}\xspace}
\newcommand{\LDCgame}{\ensuremath{\mathtt{LDC}\text{-}\mathtt{Game}}\xspace}

\newcommand{\NL}{\mbox{}\\}

\newcommand{\Cfin}{\ensuremath{C_{\fin}}\xspace}
\newcommand{\ellf}{\ensuremath{\ell_{\fin}}\xspace}
\newcommand{\rhof}{\ensuremath{\rho_{\fin}}\xspace}
\newcommand{\pf}{\ensuremath{p_{\fin}}\xspace}
\newcommand{\epsf}{\ensuremath{\eps_{\fin}}\xspace}
\newcommand{\cAf}{\ensuremath{\cA_{\fin}}\xspace}

\newcommand{\ro}{\ensuremath{\pred{H}}\xspace}

\newcommand{\compile}{\ensuremath{\mathtt{Compile}}\xspace}
\newcommand{\recover}{\ensuremath{\mathtt{RecoverBit}}\xspace}

\newcommand{\reduce}{\ensuremath{\mathtt{Reduce}}\xspace}

\title{Private and Resource-Bounded Locally Decodable Codes for Insertions and Deletions\IEEEauthorrefmark{1}\thanks{\IEEEauthorrefmark{1}Full-version of the work with the same title published at ISIT 2021, available at \url{https://doi.org/10.1109/ISIT45174.2021.9518249}. \copyright 2021 IEEE. Personal use of this material is permitted. Permission from IEEE must be obtained for all other uses, in any current or future media, including reprinting/republishing this material for advertising or promotional purposes,creating new collective works, for resale or redistribution to servers or lists, or reuse of any copyrighted component of this work in other works.}} 

\author{%
\IEEEauthorblockN{Alexander R. Block and Jeremiah Blocki}
\IEEEauthorblockA{Purdue University\\
				  Email: \texttt{\{block9, jblocki\}@purdue.edu}\IEEEauthorrefmark{2}\thanks{\IEEEauthorrefmark{2}Jeremiah Blocki was supported in part by NSF CNS \#1704587, NSF CNS \#1755708 and NSF CCF \#1910659. Alexander R. Block was supported in part by NSF CCF \#1910659.}}
}

\IEEEoverridecommandlockouts

\begin{document}




\maketitle


\begin{abstract}
	We construct locally decodable codes (LDCs) to correct insertion-deletion errors in the setting where the sender and receiver share a secret key or where the channel is resource-bounded.
	Our constructions rely on a so-called ``Hamming-to-InsDel'' compiler (Ostrovsky and Paskin-Cherniavsky, ITS '15 \& Block \etal, FSTTCS '20), which compiles any locally decodable Hamming code into a locally decodable code resilient to insertion-deletion (InsDel) errors.
	While the compilers were designed for the classical coding setting, we show that the compilers still work in a secret key or resource-bounded setting.
	Applying our results to the private key Hamming LDC of Ostrovsky, Pandey, and Sahai (ICALP '07), we obtain a private key InsDel LDC with constant rate and polylogarithmic locality.
	Applying our results to the construction of Blocki, Kulkarni, and Zhou (ITC '20), we obtain similar results for resource-bounded channels; \ie, a channel where computation is constrained by resources such as space or time.
%
%
\end{abstract}




\section{Introduction}
Error-correcting codes that are resilient to insertion-deletion (InsDel) errors have been a major focus in recent advances in coding theory~\cite{Levenshtein_SPD66,Kiwi_expectedlength,guruswami2017deletion,STOC:HaeSha17,guruswami2019polynomial,SODA:GurLi18,ICALP:HaeShaSud18,STOC:HaeSha18,ISIT:BraGurZba18,FOCS:CJLW18,SODA:CHLSW19,ICALP:CJLW19,STOC:HaeRubSha19,FOCS:Haeupler19,ISIT:SimBru19,CGHL20,CheLi20,STOC:GurHaeSha20,liu2020list}.
Such codes are a generalization of classical Hamming codes to handle the case where symbols at arbitrary positions in the codeword can be inserted or deleted.
Insertion-deletion codes over alphabet $\Sigma$ are described by an encoding function $\Enc \colon \Sigma^k \rightarrow \Sigma^K$ and decoding function $\Dec \colon \Sigma^* \rightarrow \Sigma^k$ such that for a message $x\in \Sigma^k$, if $y' \in \Sigma^*$ such that the edit distance between $\Enc(x)$ and $y'$ is at most $2\rho K$, then $\Dec(y') = x$.
A core research direction is building codes with high information rate $k/K$ that are robust to a large constant fraction $\rho$ of insertion-deletion errors.
Only recently have {\em efficient} (\ie, polynomial time encoding and decoding) InsDel codes with asymptotically good (\ie, constant) information rate and error tolerance been well-understood~\cite{STOC:HaeSha18,FOCS:Haeupler19,STOC:HaeRubSha19,liu2020list,STOC:GurHaeSha20}.

Even less understood are {\em locally decodable} codes for insertions and deletions: such error-correcting codes admit super efficient (\eg, polylogarithmic time) decoding algorithms which, by querying few locations into a received word, can recover portions of the original message.
Inspired by locally decodable codes (\LDC{}s) for Hamming errors \cite{CCC:SudTrevVad99,STOC:KatTre00}, Ostrovsky and Paskin-Cherniavsky \cite{OstPan15} introduced the notion of {\em locally decodable InsDel codes} (InsDel \LDC).
A code $C[K,k]=(\Enc,\Dec)$ is an $(\ell, \rho, p)$-InsDel \LDC if the decoding function $\Dec$ is a randomized algorithm that makes at most $\ell$ queries to the received word and, if the edit distance between an encoded message $\Enc(x)$ and a received word $y'$ is at most $2\rho\cdot|\Enc(x)|$, then $\Dec$ on input $i$ outputs $x_i$ with probability at least $p$.
Here, $\ell$ is the {\em locality} of the code, $\rho$ is the {\em error rate}, and $p$ is the {\em success probability}.
While \LDC{}s for Hamming errors have been studied for several decades~\cite{STOC:KerWol03,Yekhanin08,STOC:Efremenko09,FOCS:DviGopYek10,Yekhanin12,KopSar16,KMRS17}, the study of InsDel \LDC{}s is scarce.
Besides the results of Ostrovsky and Paskin-Cherniavsky~\cite{OstPan15} and Block \etal \cite{BBGKZ20}, 
only Haeupler and Shahrasbi \cite{STOC:HaeSha18}, to the best of our knowledge, consider locality in the building of synchronization strings, which are an important component of optimal InsDel codes.

Ostrovsky and Paskin-Cherniavsky \cite{OstPan15} and Block \etal \cite{BBGKZ20} both give a so-called ``Hamming-to-InsDel'' compiler: given any classical Hamming \LDC as input, this compiler outputs an InsDel \LDC.
This reduction preserves the information rate and the error rate of the original Hamming \LDC (up to constant factors), and the locality only grows by a polylogarithmic factor (in the length of a codeword).
Note this reduction holds for any {\em classical} Hamming \LDC.
However, there have been recent advances in examining Hamming \LDC{}s in non-classical \cite{ICALP:OstPanSah07,EPRINT:CheLiZhe20,ITC:BloKulZho20} or relaxed \cite{ISIT:BGGZ19} settings. 
For example, there is a line of work studying Hamming codes in which the channel is computationally bounded \cite{Lipton94,TCC:MPSW05,GurSmi16,ShaSil16}. 
In such settings the corruption pattern is selected adversarially by a resource bounded channel (\eg, the channel is probabilistic polynomial time), or has other resource constraints such as space or computation depth (\ie, sequential time), restricting the computations that can be performed. 
It has been argued that any real world communication channel can be reasonably modeled as a resource-bounded channel \cite{Lipton94,ITC:BloKulZho20}. 
The notion of resource-bounded channels is well-motivated by channels in the real world, which all have some sort of limitations on their computations, and one can reasonably expect error patterns encountered in nature to be modeled by some (not necessarily known) probabilistic polynomial time algorithm.
Thus, the study of Hamming and InsDel codes in non-classical and relaxed settings is well-motivated.

Mirroring the Hamming code results, the non-classical and relaxed settings offer much better tradeoffs than classical \LDC{}s, at the cost of different assumptions in the adversarial models, or by allowing the decoder to fail on a small fraction of inputs.
For example, codes constructed using secret-key cryptography (\ie, the encoder and decoder share a secret key) admit constant-rate Hamming \LDC{}s with polylogarithmic locality (in the security parameter) \cite{ICALP:OstPanSah07}.
Similarly, when assuming the adversarial channel is resource-constrained in some way (\eg, the channel is a low-depth circuit), there are constructions of constant-rate Hamming \LDC{}s with polylogarithmic locality \cite{ITC:BloKulZho20}.
Further, it is not out of the question for a shared secret-key assumption, and it has been argued that resource-constrained adversarial channels can model real-world channels reasonably well \cite{Lipton94,ITC:BloKulZho20}.
Thus we ask
\begin{center}
	{\em Can we extend non-classical Hamming \LDC{}s to the insertion-deletion setting?}
\end{center}

\subsection{Our Results}
We answer the question in the affirmative for two classes of non-classical Hamming \LDC{}s.
First, we consider {\em private locally decodable codes} (private \LDC{}s).
Introduced by Ostrovsky, Pandey, and Sahai~\cite{ICALP:OstPanSah07}, private \LDC{}s leverage cryptographic assumptions to construct locally decodable Hamming codes against probabilistic polynomial time adversaries.
In particular, private \LDC{}s leverage a (pseudorandom) {\em secret key} that is shared between the encoder and the decoder, and assumes that any adversary does not receive this secret key.
These codes are additionally parameterized by a security parameter $\lambda$ and a secret key generation function $\Gen$.
Second, we consider Hamming \LDC{}s that are secure against {\em resource-bounded adversaries}.
Blocki, Kulkarni, and Zhou~\cite{ITC:BloKulZho20} introduce resource-bounded \LDC{}s as an extension of classical Hamming codes in resource-bounded settings~\cite{Lipton94,TCC:MPSW05,GurSmi16,ShaSil16}.
These \LDC{}s are secure against any class of adversaries $\bbC$ that admit some {\em safe function} that is uncomputable by any adversary $\cA \in \bbC$. For example, in the (parallel) random oracle model any polynomial time algorithm running in sequential time $T$ provably cannot evaluate the function $\ro^{T+1}(\cdot)$ so the function would be a safe function against the class $\bbC$ of probabilistic polynomial time algorithms with sequential time at most $T$.

We obtain a binary private InsDel \LDC from any private Hamming \LDC and a binary resource-bounded InsDel \LDC from any resource-bounded Hamming \LDC by applying the Hamming-to-InsDel compiler of Block \etal~\cite{BBGKZ20}.
\begin{informalthm}[see \cref{thm:priv-insdel}]\label{infthm:priv}Let $C[K,k]$ be an $(\ell,\rho,p)$-private Hamming \LDC.
	There exists a binary $(\ell\cdot\polylog(K),\Theta(\rho),O(p))$-private InsDel \LDC with codeword length $\Theta(K)$.
\end{informalthm}

Ostrovsky, Pandey, and Sahai~\cite{ICALP:OstPanSah07} construct a private Hamming \LDC over any constant-sized alphabet that achieves constant-rate, $\omega(\log (\lambda))$ locality, constant error rate, and success probability $1-\negl(\lambda)$, where $\lambda$ is the security parameter and $\negl(\cdot) = o(1/|p(\cdot)|)$ for any non-zero polynomial $p$.
Combining \cite{ICALP:OstPanSah07} with \cref{infthm:priv} yields a constant-rate private InsDel \LDC with polylogarithmic locality, constant error rate, and high success probability.
\begin{informalthm}[see~\cref{thm:resource-insdel}]\label{infthm:resource-bounded}Let $C[K,k]$ be an $(\ell, \rho,p)$-Hamming \LDC secure against class $\bbC$.
	There exists a binary $(\ell\cdot\polylog(K),\Theta(\rho), O(p))$-InsDel \LDC secure against class $\bbC$ with codeword length $\Theta(K)$.
\end{informalthm}

Blocki, Kulkarni, and Zhou \cite{ITC:BloKulZho20} recently construct a Hamming \LDC over any constant-sized alphabet, in the random oracle model (\ie, the encoding and decoding functions make use of a cryptographic hash function), that achieves constant-rate, $\polylog(\lambda)$ locality, constant error rate, and success probability $1 - \negl(\lambda)$, for security parameter $\lambda$. In the random oracle model their construction provably yields a secure code for any channel class $\bbC$ admiting a safe function. 
Combining~\cite{ITC:BloKulZho20} with \cref{infthm:resource-bounded} yields a constant rate InsDel \LDC secure against class $\bbC$ with polylogarithmic locality, constant error rate, and high success probability.



\subsection{Technical Overview}\label{sec:overview}
The key technical component of our constructions is the use of a ``Hamming-to-InsDel'' compiler ~\cite{OstPan15,BBGKZ20} which  transforms any classical Hamming \LDC to an InsDel \LDC with polylogarithmic blow-up in the locality.
The compiler of Block \etal~\cite{BBGKZ20} is a reproving of Ostrovsky and Paskin-Cherniavsky's result, using different techniques and analysis.
For simplicity, we use the compiler of Block \etal in this work, which we refer to as the BBGKZ compiler.


The BBGKZ compiler at its core consists of two functions: $\compile$ and $\recover$.
The function $\compile$ takes as input a codeword $y \in \Sigma^K$ that is resilient to $\rho$-fraction of Hamming errors and outputs a codeword $Y \in \zo^n$ that is resilient to $\rho'$-fraction of insertion-deletion errors.
The compiled encoding function operates as follows: it encodes a message $x$ using the given Hamming \LDC to obtain the Hamming codeword $y$, then it applies the function $\compile$ to $y$ and outputs the final InsDel codeword $Y$.
The function \recover, when given query access to some $Y' \in \zo^*$, on input $i$ makes $\polylog(|Y'|)$ queries to $Y'$ and attempts to recover $y_i$, the $i$th bit of the Hamming codeword $y$. The BBGKZ compiler guarantees that if $\ED(Y,Y') \leq \rho'$ then for {\em most} indices $i \in [K]$, \recover outputs the correct bit $y_i$ with high probability.

The challenge in applying the BBGKZ compiler to a private Hamming \LDC or a resource-bounded \LDC is that we cannot assume that decoding will be correct for {\em every} corrupted codeword with small Hamming distance. 
Instead, we require that the channel cannot produce a codeword which fools the decoding algorithm except with negligible probability. 
In particular, if $y$ is our encoding of a message $x$ then we say that a corrupted codeword $y'$ {\em fools} the decoder if: 
\begin{enumerate}
	\item the (Hamming/Edit) distance between $y$ and $y'$ is small; and
	\item for some index $i$, the probability that the local decoder, given oracle access to $y'$, outputs the correct bit $x_i$ is less than $p$.
\end{enumerate}
The security requirement is that any adversary $\cA$ produces such a fooling codeword $y'$ with probability at most $\eps$.
The difficulty here is proving that applying the BBGKZ compiler to a private code or resource-bounded code preserves the security of the underlying code. 
Proving the security of our compiled private/resource-bounded code lies in the algorithm $\recover$: given an adversary $\cA$ against the compiled InsDel code, we construct a new adversary $\cA'$ against the original Hamming code which does the following: 
\begin{enumerate}
	\item obtains challenge message $x$ and Hamming codeword $y$;
	\item obtains InsDel codeword $Y = \compile(y)$;
	\item obtains $Y' \gets \cA(x,Y)$; and
	\item obtains $y_j' \gets \recover^{Y'}(j)$ for all $j$.
\end{enumerate}
Applying the key property of \recover one can show that the Hamming distance between $y$ and $y'$ is suitably small. 
Furthermore, if $Y'$ fools our local InsDel decoder then one can argue that (\whp) $y'$ fools our local Hamming decoder. 
Thus, the compiler transforms secret key Hamming LDCs into secret key InsDel LDCs and resource bounded Hamming LDCs into resource bounded InsDel LDCs. 
For resource bounded channels, there is another subtlety we must account for.
Our Hamming adversary $\cA'$ requires {\em slightly} more resources than the original InsDel adversary $\cA$; i.e., we need to run \recover for each index $j$ (though this can be accomplished in parallel to minimize computation depth). Thus, to obtain an InsDel LDC secure against the channel class $\bbC$ we need to start with a Hamming LDC secure against a slightly larger class $\bbC'$.  


\subsection{Related Work}
Levenstein~\cite{Levenshtein_SPD66} initiated the study of codes for insertions and deletions.  
Since this initiation, there has been a large body of works examining InsDel codes (see surveys \cite{Sloane2002OnSC,SWAT:Mitzenmacher08,IEEE:MerBhaTar10}).
Recently, \cite{ISIT:SimBru19} constructed $k$-deletion correcting binary codes with optimal redundancy, which was extended to systematic binary codes and $q$-ary codes in~\cite{ISIT:SimGabBru20b,ISIT:SimGabBru20a}.
This line of work answered long standing open problems in the construction of $k$-deletion correcting codes with optimal redundancy.
Random codes with positive information rate and correcting a large fraction of deletion errors were studied in \cite{Kiwi_expectedlength,guruswami2017deletion}, and efficiently encodable and decodable codes with constant rate and resilient to a constant fraction of insertion-deletion errors were studied extensively in \cite{SchZuc99,guruswami2017deletion,STOC:HaeSha17,FOCS:CJLW18,STOC:HaeSha18,SODA:CHLSW19,guruswami2019polynomial,ISIT:BraGurZba18,CGHL20,CheLi20,STOC:GurHaeSha20}.
Recently, there has been interest in extending ``list-decoding'' to the setting of InsDel codes.
These codes are resilient to a larger fraction of insertion-deletion errors at the cost of outputting a small list of potential codewords (\ie, the loss of unique decoding) \cite{ICALP:HaeShaSud18,liu2020list,STOC:GurHaeSha20}.
Another direction due to Haeupler and Shahrasbi \cite{STOC:HaeSha18} involves constructing explicit synchronization strings which can be ``locally decoded'' in the following sense: each index of the string can be computed using values located at a small number of other indices.
These explicit and locally decodable synchronization strings are used to imply near linear time interactive coding schemes for insertion-deletion errors.

Cheng, Li and Zheng \cite{EPRINT:CheLiZhe20} propose the notion of locally decodable codes with randomized encoding, in both the Hamming and edit distance regimes.
They study such codes in various settings, including where the encoder and decoder share randomness, or the channel is oblivious to the codeword, and hence adds error patterns non-adaptively. 
For insertion-deletion errors they obtain codes with $K=O(k)$ or $K= k \cdot \log(k)$ and $\polylog(k)$ locality for message length $k$.

Blocki, Gandikota, Grigorescu, and Zhou \cite{ISIT:BGGZ19} construct {\em relaxed} locally correctable and locally decodable Hamming codes in computationally bounded channels.
Here, {\em local correction} states that a corrupt codeword $c'$ can be corrected to codeword $c$ by only querying $c'$ at a bounded number of locations, and {\em relaxed} means that the correcting or decoding algorithm is allowed to output the value $\bot$ for a small fraction of inputs.
Their construction requires a public parameter setup for a collision-resistant hash function, and they obtain relaxed binary locally correctable and decodable Hamming codes with constant information rate and polylogarithmic locality.
Recently, Blocki, Kulkarni, and Zhou~\cite{ITC:BloKulZho20} introduced Hamming \LDC{}s that are secure against resource-bounded adversaries, in the random oracle model.
Here, they construct codes (in the random oracle model) which are resilient to classes of adversaries $\bbC$ for which there exists a function $f$ that is uncomputable by any $\cA \in \bbC$.
They obtain explicit  Hamming \LDC{}s with constant information rate and polylogarithmic locality against various classes $\bbC$ of resource-bounded adversaries. 



\section{Preliminaries}\label{sec:prelims}
We let $\lambda \in \bbN$ denote the security parameter.
For $n \in \bbZ^+$, we let $[n]$ denote the set $\{1,2,\dotsc, n\}$.
A function $\vartheta \colon \bbN \rightarrow \bbR_{\geq 0}$ is said to be negligible if $\vartheta(n) = o(1/|p(n)|)$ for any fixed non-zero polynomial $p$.
We write PPT as a shorthand for probabilistic polynomial time.
For any (randomized) algorithm $A$, we let $y \gets A(x)$ denote the result of running $A$ on some input $x$.

We consider the {\em fractional Hamming distance} and the {\em fractional Edit Distance} metrics, which we denote by \HAM and \ED, respectively.
For two strings $x,y \in \Sigma^K$ for some $K$, we define $\HAM(x,y) \defeq |\{ i \colon x_i \neq y_i \}_{i \in [K]}|/K$.
For two strings $x \in \Sigma^K$ and $y \in \Sigma^*$, we define $\ED(x,y)$ is the minimum number of insertions and deletions required to transform $x$ into $y$ (or vice versa), normalized by $2K$.

\begin{definition}[Error-correcting Codes]\label{def:ecc}A {\em coding scheme} $C[K,k,q_1,q_2] = (\Enc, \Dec)$ is a pair of encoding and decoding algorithms $\Enc\colon \Sigma_1^k \rightarrow \Sigma_2^K$ and $\Dec \colon \Sigma_2^* \rightarrow \Sigma_1^k$, where $|\Sigma_i| = q_i$.
	A code $C[K,k,q_1,q_2]$ is a {\em $(\rho,\dist)$ error-correcting code} for $\rho \in [0,1]$ and fractional distance $\dist$ if for all $x \in \Sigma_1^k$ and $y \in \Sigma_2^*$ such that $\dist(\Enc(x),y) \leq \rho$, we have that $\Dec(y) = x$.
	Here, $\rho$ is the {\em error rate} of $C$.
	If $q_1 = q_2$, we simply denote this by $C[K,k,q_1]$.
	If $\dist = \HAM$, then $C$ is a {\em Hamming code}; if $\dist = \ED$, then $C$ is an {\em insertion-deletion code} (InsDel code).
\end{definition}

\begin{definition}[Locally Decodable Codes]\label{def:ldc}A coding scheme $C[K,k,q_1,q_2] = (\Enc,\Dec)$ is an {\em $(\ell, \rho, p, \dist)$-locally decodable code} (\LDC) if for all $x \in \Sigma_1^k$ and $y \in \Sigma_2^*$ such that $\dist(\Enc(x), y) \leq \rho$, the algorithm $\Dec$, with query access to word $y$, on input index $i \in [k]$, makes at most $\ell$ queries to $y$ and outputs $x_i$ with probability at least $p$ over the randomness of the decoder.
	Here, $\ell$ is the {\em locality} of $C$ and $p$ is the {\em success probability}.
\end{definition}


Private locally-decodable codes were introduced by Ostrovsky, Pandey, and Sahai \cite{ICALP:OstPanSah07}.
The encoding and decoding algorithms of these codes additionally share a secret key that is hidden from any adversarial channel. Intuitively, these codes ensure that (except with small probability) any channel who does not have the secret key will fail to produce a corrupted codeword $y'$ which fools the local decoder.
\begin{definition}[One-Time Private LDC]\label{def:priv-ldc}Let $\lambda$ be the security parameter.
	A code $C[K,k,q_1,q_2,\lambda]$ consisting of a tuple of PPT algorithms $(\Gen,\Enc,\Dec)$ is a {\em $(\ell,\rho,p,\eps,\dist)$-one time private locally decodable code (private \LDC)} if: 
	\begin{itemize}
		\item $\Gen(1^\lambda)$ is the key generation algorithm that takes $1^\lambda$ as input and outputs a secret key $\sk \in \zo^*$, for security parameter $\lambda$;
		\item $\Enc\colon \Sigma_1^{k} \times \zo^* \rightarrow \Sigma_2^K$ is the encoding algorithm that takes as input a message $x \in \Sigma_1^k$ and a secret key $\sk$ and outputs a codeword $y \in \Sigma_2^K$; and
		\item $\Dec^{y'} \colon \zo^{\log k} \times \zo^* \rightarrow \Sigma_1$ is the decoding algorithm that takes as input index $i\in [k]$ and secret key $\sk$, and is additionally given query access to a corrupted codeword $y' \in \Sigma_2^{K'}$ and outputs $b \in \Sigma_1$ after making at most $\ell$ queries to $y'$. 
	\end{itemize}
	We define a predicate $\pred{Fool}(y',\rho,p, \sk,x, y) = 1$ if and only if
	\begin{enumerate}
		\item $\dist(y, y') \leq \rho$; and
		\item $\exists i \in [k]$ such that $\Pr[\Dec^{y'}(i,\sk)=x_i] < p$, where the probability is taken over the random coins of $\Dec$.
	\end{enumerate}
	We require that for all adversaries $\cA$ and all $x \in \Sigma_1^k$, 
	\begin{align*}
		\Pr[ \pred{Fool}(\cA(y), \rho, p, \sk, x, y) = 1 ] \leq \eps,
	\end{align*}
	where $y \leftarrow \Enc(x,\sk)$ and the probability is taken over the random coins of $\cA$ and $\Gen$ and $\Enc$  (if encoding is randomized).
\end{definition}

For all of our code definitions, when $q_2 = 2$ 
we say that the code is a {\em binary code}.

\subsection{Codes for Resource-Bounded Channels}
Recently, Blocki, Kulkarni, and Zhou \cite{ITC:BloKulZho20} studied error-correcting codes against channels which have some resource bound; e.g., the channel is a low-depth circuit, or is a one-tape Turing machine.
Intuitively, these codes ensure that (except with small probability) any adversary with insufficient resources will fail to produce a corrupt codeword $y'$ which fools the local decoder.
\begin{definition}[$\bbC$-secure LDC]\label{def:c-secure}A code $C[K,k,q_1,q_2] = (\Enc, \Dec)$ is a {\em$(\ell, \rho, p, \eps, \dist, \bbC)$-locally decodable code against class $\bbC$} if $\Dec$ takes as input index $i \in [k]$, is additionally given query access to a corrupted codeword $y' \in \Sigma_2^{K'}$, and outputs $b \in \Sigma_1$ after making at most $\ell$ queries to $y'$.
	We define predicate $\pred{Fool}(y', \rho, p, x, y) = 1$ if and only if
	\begin{enumerate}
		\item $\dist(y,y') \leq \rho$; and
		\item $\exists i \in [k]$ such that $\Pr[\Dec^{y'}(i) = x_1] < p$,
	\end{enumerate} 
	where the probability is taken over the random coins of $\Dec$; otherwise $\pred{Fool}(y', \rho, p, x, y) = 0$.
	We require that for all adversaries $\cA \in \bbC$ and all $x \in \Sigma_1^k$, 
	\begin{align*}
		\Pr[\pred{Fool}(\cA(y), \rho,p,y) = 1] \leq \eps,
	\end{align*}
	where the probability is taken over the random coins of $\cA$ and the generation of the codeword $y \leftarrow \Enc(x)$. 
\end{definition}

\subsection{Hamming-to-InsDel Compiler}
Ostrovsky and Paskin-Cherniavsky~\cite{OstPan15} give a compiler which transforms any Hamming \LDC to an InsDel \LDC with a polylogarithmic blowup in locality.
Block \etal \cite{BBGKZ20} give another compiler which transforms any Hamming \LDC into an InsDel \LDC with polylogarithmic blow-up in locality, reproving the result of~\cite{OstPan15} with different techniques and analysis.
We use the compiler of Block \etal in this work.
%

Let $C = (\Enc, \Dec)$ be a Hamming \LDC.
Then the compiler works as follows.
The compiled encoder is defined as $\Encf(x) \defeq \compile(\Enc(x))$ for any message $x$.
The decoder $\Decf$ contains a subroutine $\recover$ which, given query access to some $Y' \in \zo^{n'}$, on input index $i$ makes at most $O(\log^4(n'))$ queries and with high probability recovers the $i$\th-bit of $c$ correctly for {\em most} indices of $c=\Enc(x)$ as long as $\ED(Y,Y')$ is sufficiently small.
The decoder $\Decf$ then runs $\Dec$ and simulates oracle access to $c$ by using algorithm $\recover$.
We formally capture the properties of the compiler in the following lemma.
\begin{lemma}[Block \etal~\cite{BBGKZ20}]\label{lem:bbgkz}There exist functions $\compile$ and $\recover$ such that for any constant $\rho > 0$ and any Hamming \LDC $C[K,k,q_1,q_2] = (\Enc, \Dec)$ with locality $\ell$, there exists $\rhof = \Theta(\rho)$ such that for any message $x$ and any $c'$ with $\ED( c', y ) \leq \rhof$ for $y = \compile(\Enc(x)) \in \zo^*$: 
	\begin{enumerate}
		\item $\Decf$ has locality $\ell\cdot O(\log^4(K\cdot\log(q_2)))$ and $|y| = \Theta(K\cdot\log(q_2))$;
		\item For $c'' = \recover^{Y'}(1)\circ \cdots \circ \recover^{Y'}(K\cdot\log (q_2))$, we have
		$\Pr[\Decf^{c'}(i) = x_i] \geq \Pr[ \Dec^{c''}(i) = x_i ] - \vartheta_1(K\cdot\log(q_2))$; and 
		\item if $\ED(c', \Encf(x)) \leq \rhof$ then, except with probability $\vartheta_2(K\cdot \log(q_2))$, $\HAM(c'', \Enc(x)) \leq \rho$.
	\end{enumerate}
	Here, $\vartheta_1$ and $\vartheta_2$ are fixed negligible functions, $\compile$ is computable in parallel time $\polylog(K)$, and $c''$ is computable in parallel time $\polylog(K)$.
\end{lemma}



\section{One-Time Private Locally Decodable Codes for Insertion-Deletion Channels}
\begin{theorem}\label{thm:priv-insdel}Let $C[K,k,q_1,q_2,\lambda]$ be a $(\ell, \rho, p, \eps, \HAM)$-one time private Hamming \LDC for constants $\rho, p > 0$.
	There exists a binary code $C_{\fin}[n,k,q_1,2]$ that is a $(\ellf, \rhof, \pf, \epsf, \ED)$-one time private InsDel \LDC, where 
	$\ellf = \ell \cdot O(\log^4(n))$, $\rhof = \Theta(\rho)$, $\pf < p$, $\epsf =\eps/(1-(\pf/p)-(\vartheta_1(n)/p)-\vartheta_2(n))$, and $n = \Theta(K \cdot \log(q_2))$.
	Here, $\vartheta_1, \vartheta_2$ are fixed negligible functions.
\end{theorem}
\begin{IEEEproof}
	Let $C[K,k,q_1,q_2, \lambda] = (\Gen, \Enc, \Dec)$ be a $(\ell,\rho,p,\eps,\HAM)$-one time private Hamming \LDC.
	We define $\Genf(1^\lambda) \defeq \Gen(1^\lambda)$.
	Then for any message $x$ and secret key $\sk$ we define $\Encf(x, \sk) \defeq \compile(\Enc(x,\sk))$.
	Fixing the secret key $\sk$ and applying \cref{lem:bbgkz} to the encoding scheme, we see that $\Decf$ has locality $\ell \cdot O(\log^4(n))$ and the output length of $\Encf$ is $n = \Theta(K \log q_2)$ bits.
	The main challenge is proving the security.
	Suppose towards contradiction that there exists an adversary $\cAf$ such that $\Pr[\pred{Fool}(\cAf(Y), \rhof, \pf, \sk, x, Y)=1] > \epsf$ for $Y \gets \Encf(x,\sk)$.
	Then we construct an adversary $\cA$ such that $\Pr[\pred{Fool}(\cA(y), \rho, p, \sk, x, y) = 1] > \eps$ for $y \gets \Enc(x,\sk)$. 
	Adversary $\cA$ works as follows:
	\begin{enumerate}
		\item $\cA$ obtains as input $x, y, \lambda, \rho, p, k,$ and $K$, where $y = \Enc(x, \sk)$;
		\item $\cA$ then obtains $Y = \compile(y)$; and
		\item $\cA$ then obtains $Y' \gets \cAf(x,Y,\lambda, \rhof, \pf, k, n )$.
	\end{enumerate}
	By assumption $\ED(Y,Y') \leq \rhof$ and with probability at least $\epsf$ there exists index $i \in [k]$ such that
	\begin{align*}
		\Pr[\Decf^{Y'}(i,\sk) = x_i] < \pf.
	\end{align*}
	$\cA$ then outputs word 
	\begin{align*}
		y' = \recover^{Y'}(1)\circ \cdots \circ \recover^{Y'}(K\cdot\log(q_2)).
	\end{align*}
	
	Suppose that $\pred{Fool}(Y', \rhof, \pf, \sk, x, Y) = 1$.
	Then we have that $\ED(Y,Y') \leq \rhof$ and there exists $i \in [k]$ such that
	\begin{align*}
		\Pr[ \Decf^{Y'}(i, \sk) = x_i ] < \pf.
	\end{align*}
	By \cref{lem:bbgkz}, we have that $\HAM(y,y') \leq \rho$ with probability at least $1 - \vartheta_2(n)$. 
	By definition of $\Decf$ and \cref{lem:bbgkz}, we have that 
	\begin{align*}
		\pf &> \Pr[ \Decf^{Y'}(i, \sk) = x_i  ]\\
		&\geq \Pr[ \Dec^{y'}(i,\sk) = x_i ] - \vartheta_1(n),
	\end{align*}
	where the randomness of the second term is taken over the coins of $\Dec$ and the coins used by $\recover$ to generate $y'$, and the randomness of the first term is taken only over the coins of $\Decf$. 
	Define the predicate $B_{p}(y')=1$ if and only if $\Pr[ \Dec^{y'}(i,\sk) = x_i ] < p$, and $B_p(y') = 0$ otherwise, where the probability is taken over $\Dec$'s coins.
	Let $\alpha = \Pr[ B_{p}(y')]$, where the probability is taken over the random coins used to generate $y'$ from $Y'$.
	Then we have that
	\begin{align*}
		\Pr[ \Dec^{y'}(i,\sk) = x_i ] \geq p(1-\alpha).
	\end{align*}
	This implies that  $\alpha > 1 - (\pf/p) - (\vartheta_1(n)/p)$. 
	Now consider two events $\cF_{\HAM} = \pred{Fool}(y',\rho,p,\sk,x,y)$ and $\cF_{\ED} = \pred{Fool}(Y', \rhof, \pf, \sk, x, Y)$.
	Then
	\begin{align*}
		\Pr[\cF_{\HAM} = 1] \geq \Pr[\cF_{\ED} = 1] \cdot \Pr[\cF_{\HAM} = 1 | \cF_{\ED} = 1].
	\end{align*}
	By assumption we have that $\Pr[\cF_{\ED} = 1] > \epsf$.
	Further, by \cref{def:priv-ldc}, $\cF_{\HAM} = 1$ if and only if $\HAM(y,y') \leq \rho$ and there exists $i \in [k]$ such that $\Pr[ \Dec^{y'}(i,\sk) = x_i ] < p$.
	Since $\cF_{\ED} = 1$, we have that $\ED(Y,Y') \leq \rhof$, and thus by \cref{lem:bbgkz} we have that $\HAM(y,y') \leq \rho$ with probability at least $1-\vartheta_2(n)$. 
	Thus
	\begin{align*}
		\Pr[\cF_{\HAM} = 1 | \cF_{\ED} = 1] \geq 1-\vartheta_2(n) - (1- \alpha)
	\end{align*} and $\alpha > 1 - (\pf/p) - (\vartheta_1(n)/p)$.
	Therefore we have that
	\begin{align*}
		\Pr[ \cF_{\HAM} = 1 ] &> \epsf \cdot (1 - (\pf/p) - (\vartheta_1(n)/p) - \vartheta_2(n)),
	\end{align*} 
	which is a contradiction since the right hand side of the above equation is equal to $\eps$.
\end{IEEEproof}

\section{Locally Decodable Codes for Resource-Bounded Insertion-Deletion Channels}
To construct \LDC{}s for resource-bounded InsDel channels, we first need to introduce the notion of {\em closure} between algorithms classes.
Let $\bbC$ be a class of parallel algorithms running in at most sequential time $T$ and maximum space usage $S$.
For any $A \in \bbC$, let $B = \reduce(A)$ be a reduction from algorithm $A$ to $B$.
We say the class of algorithms $\bbC'$ is the {\em closure of \bbC with respect to \reduce} if $\bbC'$ is the minimum class of algorithms such that $\reduce(A) \in \bbC'$ for all $A \in \bbC$.

In our context, for parameter $N$ we define $\reduce_N$ as a sequential time $N \cdot\polylog(N)$ reduction that can be executed in parallel for sequential time $\polylog(N)$.
Parallel execution incurs an additional $N \cdot\polylog(N)$ space overhead, and sequential execution incurs an additional $\polylog(N)$ space overhead.
Thus, if $\bbC$ is the class of all parallel PPT algorithms running in sequential time $T$, then $\overline{\bbC}(N)$ is some class of parallel PPT algorithms running in time $T + \polylog(N)$.

\begin{theorem}\label{thm:resource-insdel}Let $\bbC$ be the class of parallel PPT algorithms running in sequential time $T$ and space $S$, and let $C[K,k,q_1,q_2] = (\Enc, \Dec)$ be a $(\ell,\rho,p,\eps,\HAM,\overline{\bbC}(n))$-\LDC for constants $\rho, p > 0$ and $n = O(K \cdot \log(q_2))$.
	There exists a binary code $\Cfin[n,k,q_1,2]$ that is a $(\ellf, \rhof, \pf, \epsf,\ED,\bbC)$-\LDC against class $\bbC$, where $\ellf = \ell\cdot O(\log^4(n))$, $\rhof = \Theta(\rho)$, $\pf < p$, and $\epsf = \eps/(1-(\pf/p) - (\vartheta_1(n)/p) -\vartheta_2(n))$.
	Here, $\vartheta_1, \vartheta_2$ are fixed negligible functions.
\end{theorem}
\begin{IEEEproof}
	The proof follows nearly identically to the proof of \cref{thm:priv-insdel}; namely, we obtain $\Cfin$ in an identical manner by using the compiler of \cref{lem:bbgkz} with the code $C$ defined above.
	The main challenge again is the security proof: given adversary $\cAf \in \bbC$ such that $\Pr[\pred{Fool}(\cA(Y), \rhof, \pf, x, Y)=1] > \epsf$ for $Y \gets \Encf(x)$, we construct an adversary $\cA \in \overline{\bbC}(n)$ such that $\Pr[\pred{Fool}(\cA(y), \rho, p, x, y) = 1] > \eps$ for $y \gets \Enc(x)$.
	Adversary $\cA$ is constructed identically as in the proof of \cref{thm:priv-insdel}, except now the constructed adversary only yields a contradiction if we can show that $\cA \in \overline{\bbC}(n)$.
	By \cref{lem:bbgkz}, we have that $\compile$ is a $\polylog(K) = \polylog(n)$ parallel time algorithm, and $y' = \recover^{Y'}(1)\circ \cdots \circ \recover^{Y'}(K\log q_2)$ is computable in $\polylog(n)$ parallel time.
	Finally, $\compile$ and $\recover$ are run independent of the adversary $\cAf$, we have that the total parallel time of $\cA$ is $T + \polylog(n)$, which implies $\cA \in \overline{\bbC}(n)$, yielding our contradiction.	
\end{IEEEproof}
\begin{remark}
	We focus on a simple reduction, but $\reduce$ can be defined in various different ways, so long as for any $\cAf \in \bbC$, it holds that constructed adversary $\cA \in \overline{\bbC}$.
\end{remark}

\section{Explicit Constructions}
As an application of our main results, we give two explicit constructions.

\subsection{Private InsDel Locally Decodable Code Construc}
First, we use \cref{thm:priv-insdel} with the one-time private Hamming \LDC of Ostrovsky, Pandey, and Sahai \cite{ICALP:OstPanSah07}.
For security parameter $\lambda$ and fixed negligible functions $\vartheta_1, \vartheta_2$, their code has constant-rate, locality $\omega(\log (\lambda))$, constant error-rate, success probability $1-\vartheta_1(\lambda)$, and security $\eps = \vartheta_2(\lambda)$.
\begin{corollary}\label{cor:priv-insdel} Let $\ellf \defeq \ellf(\lambda,n) = \omega(\log (\lambda) \cdot O(\log^4(n))$. There exists a binary code $\Cfin[n,k,q_1,2,\lambda]$ that is a $( \ellf, \rhof, \pf, \epsf )$-one time private InsDel \LDC with constant information rate $k/n=\Theta(1)$, where , $\rhof = \Theta(1)$, $\pf = \Theta(1)$, and $\epsf \leq \varsigma(\lambda,n)$. 	Here, $\varsigma$ is a fixed negligible function.
\end{corollary}

Both the OPS one-time private Hamming \LDC and our constructed one-time private InsDel \LDC are secure against information theoretic adversaries, so long as the secret key is picked uniformly at random.
However, it is possible to also pick the secret key in a {\em psuedo-random} manner and obtain security against any class of PPT adversaries, assuming the existence of one-way functions.

Ovstrovsky, Pandey, and Sahai also give a construction of a private locally decodable code that is secure even when the adversary is given access to {\em polynomially-many} (in the security parameter) codewords (\ie, it is not one-time).
The construction relies on a family of {\em psuedo-random functions} and is therefore secure against any class of PPT adversaries, assuming the existence of one-way functions.
We emphasize that applying our compiler on this ``multi-time'' private Hamming code yields a secure ``multi-time'' private InsDel code. 
\begin{definition}\label{def:multi-ldc}Let $\lambda$ be the security parameter.
	A code $C[K,k,q_1,q_2,\lambda]$ consisting of a tuple of PPT algorithms $(\Gen, \Enc, \Dec)$ is a {\em $(\ell, \rho, p, \dist)$-private locally decodable code} if:
	\begin{itemize}
		\item $\Gen(1^\lambda)$ is the key generation algorithm that takes $1^\lambda$ as input and outputs a secret key $\sk \in \zo^*$, for security parameter $\lambda$;
		
		\item $\Enc\colon \Sigma_1^{k} \times \zo^* \rightarrow \Sigma_2^K$ is the encoding algorithm that takes as input a message $x \in \Sigma_1^k$ and a secret key $\sk$ and outputs a codeword $y \in \Sigma_2^K$; and
		
		\item $\Dec^{y'} \colon \zo^{\log k} \times \zo^* \rightarrow \Sigma_1$ is the decoding algorithm that takes as input index $i\in [k]$ and secret key $\sk$, and is additionally given query access to a corrupted codeword $y' \in \Sigma_2^{K'}$ and outputs $b \in \Sigma_1$ after making at most $\ell$ queries to $y'$. 
		
		\item Let $\pred{Fool}$ be a predicate such that $\pred{Fool}(y',\rho,p, \sk,x, y) = 1$ if and only if \begin{enumerate*}
			\item $\dist(y,y') \leq \rho$; and
			\item $\exists i \in [k]$ such that $\Pr[\Dec^{y'}(i,\sk)=x_i] < p$, where the probability is taken over the random coins of $\Dec$;
		\end{enumerate*}
		and $\pred{Fool}(y',\rho,p, \sk,x, y) = 0$ otherwise.
		Consider the $\privLDCgame$ defined in \cref{fig:priv-ldc-game}. We require that for all PPT adversaries $\cA$ there exists a negligible function $\eps(\cdot)$ such that
		\begin{align*}
			\Pr[\privLDCgame(\cA, C, 1^\lambda, \pred{Fool}) = 1] \leq \eps(\lambda).
		\end{align*}
	\end{itemize}
\end{definition}

\begin{figure}[!h]
	\begin{boxedalgo}
		$\privLDCgame(\cA, C, 1^\lambda, \pred{Fool})\colon$\NL
		{\bfseries Input:} A PPT adversary $\cA$, a locally decodable code $C[K,k,q_1,q_2, \lambda]$ with PPT algorithms $(\Gen, \Enc, \Dec)$, security parameter $1^\lambda$, and predicate $\pred{Fool}$.
		\begin{enumerate}
			\item Obtain $\sk \gets \Gen(1^\lambda)$ and share $\sk$ with $\Enc$ and $\Dec$.
			Note $\cA$ is not given $\sk$.
			\item For $i \in [h]$, where $h = \poly(k)$ is an integer:
			\begin{enumerate}
				\item $x_i \gets \cA(1^\lambda, (x_1,y_1),\dotsc (x_{i-1}, y_{i-1}))$.
				\item $y_i \gets \Enc(x_i; \sk)$;
				\item $y_i' \gets \cA(1^\lambda, (x_1,y_1),\dotsc (x_{i}, y_i))$.
			\end{enumerate}
		\end{enumerate}
		The output of $\privLDCgame$ is $1$ if there exists $i \in [h]$ such that $\pred{Fool}(y_i', \rho, p, \sk, x_i, y_i) = 1$; else $\privLDCgame$ outputs $0$.
	\end{boxedalgo}
	\caption{Definition of \privLDCgame.}\label{fig:priv-ldc-game}
\end{figure}
\begin{remark}
	\cref{def:multi-ldc} differs slightly the original definition proposed in Ovstrovsky, Pandey, and Sahai \cite{ICALP:OstPanSah07} in that we allow the attacker to output a corrupted codeword $y_i'$ in every round $i \leq h$, while in \cite{ICALP:OstPanSah07} the attacker only attempts to corrupt the codeword in the last round. However, the two definitions are equivalent, up to a $1/\poly(\lambda)$ loss in the security. In particular, an attacker that breaks \cref{def:multi-ldc} can efficiently be transformed into an attacker that breaks the definition from \cite{ICALP:OstPanSah07} i.e., we simply guess the index $i' \leq h$ of the first round in which the attacker is successful. 
\end{remark}

For security parameter $\lambda$, the ``multi-time'' private Hamming code of \cite{ICALP:OstPanSah07} has constant rate, locality $\omega(\log^2(\lambda))$, constant error-rate, and success probability $1-\negl(\lambda)$, for some negligible function.
\begin{corollary}\label{cor:multi-priv-insdel}Assume that one-way functions exist and let $\ellf \defeq \ellf(\lambda,n)= \omega(\log^2(\lambda) \cdot O(\log^4(n))$. There exists a binary code $\Cfin[n,k,q_1,2,\lambda]$ that is a $( \ellf, \rhof, \pf )$-private InsDel \LDC (as per \cref{def:multi-ldc}) with constant information rate $k/n = \Theta(1)$, where $\rhof = \Theta(1)$, and $\pf = \Theta(1)$. 
\end{corollary}

\begin{remark}
	The reduction for \cref{cor:multi-priv-insdel} is nearly identical to that of \cref{thm:priv-insdel}, except we must now account for $\poly(k)$ rounds where the adversary attempts to fool the decoder. An identical argument uses \cref{lem:bbgkz} to that the probability of succeeding in each individual round of \privLDCgame is negligible. Since there are only polynomially many rounds the the probability that the attacker succeeds in any of the rounds is still negligible.   
\end{remark}

\subsection{Resoure-Bounded InsDel Locally Decodable Code Construction}
Next we use \cref{thm:resource-insdel} with the resource-bounded Hamming \LDC of Blocki, Kulkarni, and Zhou~\cite{ITC:BloKulZho20} which works for {\em any} class $\bbC$ that admits a {\em safe function}. A function $f \colon \zo^n \rightarrow \zo^*$ is {\em $\delta$-safe} for a class $\bbC$ of algorithms if for all $\cA \in \bbC$ we have $\Pr[\cA(x) = f(x)] \leq \delta$, where the probability is taken over the random coins of $\cA$ and the selection of an input $x \in \zo^n$. 
The code construction of \cite{ITC:BloKulZho20} is in the (parallel) {\em random oracle model}, where the encoder and decoder additionally have access to some random oracle $\ro$. 
For security parameter $\lambda$ and fixed negligible functions $\vartheta_1, \vartheta_2$, their code has constant-rate, locality $\polylog(\lambda)$, constant error-rate, success probability $1 - \vartheta_1(\lambda)$, and security $\eps \leq \vartheta_2(\lambda) + q\cdot \delta$, where $q$ is an upper bound on the number of oracle queries made by any algorithm in $\bbC$.

In the (parallel) random oracle model one can provably establish the existence of safe functions for many natural classes of channels; e.g., space bounded or sequential time bounded.  
As an example, if $\ro\colon\{0,1\}^*\rightarrow \{0,1\}^\lambda$ is a random oracle then the function $\ro^{T+1}(\cdot)$ is $\delta=q\cdot T\cdot 2^{-\lambda}$-safe against the class of algorithms making at most $q$ total queries to $\ro$ over {\em at most} $T$ rounds. 
Similar results holds for the classes of space-bounded or space-time bounded channels. 
The class of sequentially bounded channels is motivated by the observation that the depth of computation that the channel performs is restricted in most natural settings; e.g., traveling at the speed of light, it would take between 3 and 22 minutes for a transmission from Mars to reach Earth (the exact time would depend on the current orbital location of the planets). 
%
\begin{corollary}
	Let $\lambda$ be a security parameter, let $\bbC$ be a class of algorithms in the parallel random oracle model admitting a $\delta$-safe function, and let $k = \poly(\lambda)$.
	For random oracle $\ro$ there exists a binary code $\Cfin^\ro[n,k,2]$ that is a $(\ellf, \rhof, \pf, \epsf, \bbC)$-InsDel \LDC against class $\bbC$, where $\ellf = \polylog(\lambda) \cdot \log^4(n')$, $\rhof = \Theta(1)$, $\pf = \Theta(1)$, and $\epsf \leq \varsigma(\lambda,n') - q\cdot\delta$.
	Here, $q$ is an upper bound on the total queries any algorithm in $\bbC$ makes to $\ro$, $\varsigma$ is a fixed negligible function, and $n'$ is the length of a word received by the decoder.
\end{corollary}
\begin{remark}
	While the construction of~\cite{ITC:BloKulZho20} relies on the random oracle model we stress that this dependence is not inherent to our results. 
	Given any standard model construction of a Hamming \LDC for resource bounded channels we could similarly obtain a standard model InsDel \LDC for resource bounded channels by applying \cref{thm:resource-insdel}. 
	Thus, it is plausible that one could replace the random oracle model assumption with, for example, the assumption that time-lock puzzles~\cite{rivest1996time,C:BonNao00,JC:GMPY11,C:MahMorVad11,ITCS:BGJPVW16} exist. 
	
	In particular, Blocki, Kulkarni, and Zhou \cite{ITC:BloKulZho20} provide serveral examples of safe functions in various models to construct resource-bounded Hamming \LDC{}s.
	These include safe functions secure in the parallel random oracle model, safe functions which are secure against sequential time-bounded adversaries, and safe functions based on graphs with sufficiently large pebbling costs.
\end{remark}

\newpage
\bibliographystyle{alphaurl}
\bibliography{abbrev3,crypto,local}



\end{document}